\documentstyle[epsfig,12pt]{article}

\begin{document}
\title{
Phase diagram of $d=4$ Ising Model with two couplings
}
\author{ J.~L.~Alonso$^{a}$,
J.~M.~Carmona$^{a}$,\\
J.~Clemente Gallardo$^{a}$,
L.~A.~Fern\'andez$^{b}$,\\
D.~I\~niguez$^{a}$,
A.~Taranc\'on$^{a}$,
C.~L.~Ullod$^{a}$ \\}
\bigskip
\maketitle

\begin{center}
{\it a)}  Departamento de F\'{\i}sica Te\'orica, Facultad de Ciencias,\\
Universidad de Zaragoza, 50009 Zaragoza, Spain \\
{\it b)} Departamento de F\'{\i}sica Te\'orica I, Facultad de Ciencias
F\'{\i}sicas,\\
Universidad Complutense de Madrid, 28040 Madrid, Spain \\
\end{center}
\bigskip
\begin{abstract}
We study the phase diagram of the four dimensional Ising model with first 
and second neighbour couplings, specially in the antiferromagnetic region,
by using Mean Field and Monte Carlo methods. 
From the later, all the transition lines seem to be first order except that 
between ferromagnetic and disordered phases in a region including the 
first-neighbour Ising transition point. 
\end{abstract}

\newpage
Antiferromagnetism has been considered in a great variety of models in order
to find properties not present in the purely ferromagnetic (FM) system.
Mostly they have been models in two and three dimensions. For instance, in 
several works on High Temperature Superconductivity the transition from 
paramagnetic (PM) to non-pure FM ordered phases has been 
studied \cite{PLUM,KAWA,MORI,ISING2d}.
On the other hand, in 
relation to the finite temperature phase transition in pure gauge $SU(3)$,
the $d=3$ three state Potts model with negative second neighbour couplings
was unsuccessfully considered to find a new critical behaviour 
\cite{POTTS,IECSEC}.

In four dimensions, in diluted systems recently new critical exponents have
been obtained \cite{PAJJ}.
From a quantum field theory point of view, no argument appears to prevent the 
existence of a non-trivial ultraviolet
limit in an antiferromagnetic (AF) lattice $\phi^4$ theory \cite{CALLAPETR}.
In fact, Gallavotti and Rivasseau more than ten years ago considered the 
possibility of an AF action which could change the ultraviolet limit of the
pure $\phi^4$ model \cite{GALLA}.

Motivated in part by that, we decided to study the phase diagram and possible
critical behaviour of a $d=4$ Ising model with an AF phase non 
trivially equivalent to the standard FM one. As we will see, 
we have found a rich phase diagram with several phase transitions but 
none of them seems to show a new critical behaviour.


The naive way to introduce antiferromagnetism in the Ising model is to 
consider a negative
coupling. In this case the state with minimal energy for large $\beta$ is
a staggered vacuum. On a hypercubic lattice, we denote
the coordinates of site $\vec n$ as
$(n_x,n_y,n_z,n_t)$. If we make the transformation
\begin{equation}
\sigma(n_x,n_y,n_z,n_t)\to(-1)^{n_x+n_y+n_z+n_t}\sigma(n_x,n_y,n_z,n_t),
\label{TRANS}
\end{equation}
the system with negative $\beta$ is mapped onto  the positive $\beta$ one, 
both regions being exactly equivalent.

Therefore, to consider true antiferromagnetism  we must take into account 
either different geometries or more couplings, in order to make the
transformation (\ref{TRANS}) not an exact mapping. The option we have chosen
here is to add a coupling between points at a distance of $\sqrt{2}$ lattice
units.

So, we will work with an Ising model with Hamiltonian 
\begin{equation}
H=\beta_1\sum_{\vec n,\mu}\sigma(\vec n)\sigma(\vec n+\vec \mu)+
  \beta_2\sum_{\vec n,\mu,\pm \nu(\mu<\nu)}
        \sigma(\vec n)\sigma(\vec n+\vec \mu+\vec \nu),
\label{HAMILTON}
\end{equation}
on a four dimensional hypercubic lattice, with side $L$ and periodic boundary
conditions. Here $\vec \mu$ denotes the unitary
vector in the $\mu$ direction.

The transformation (\ref{TRANS})
maps the semiplane $\beta_1>0$ onto the $\beta_1<0$ one, and therefore only the
region with $\beta_1\geq0$ will be considered. On the line $\beta_1=0$ the
system decouples into two independent sublattices.

The presence of two couplings with opposite signs makes frustration
to appear, and very different vacua are possible. For
small values of $\beta_1$ and $\beta_2$ the system is disordered (PM 
phase). On the other hand, we have computed the configurations which minimize
the energy for several asymptotic values of the parameters. We have only
considered configurations with periodicity two. More complex structures have
not been observed in our simulations.

We have found the following regions:
\begin{enumerate}
\item{} Small absolute values of $\beta_1,\beta_2$. PM phase. 
\item{} Large positive $\beta_1,\beta_2$, or large $\beta_1>0, \beta_2<0 $ with
 $\beta_1>6\vert \beta_2 \vert$.
 The vacuum is the FM one,
$\sigma(\vec n)=\sigma_0$ ($\sigma_0$ stands for a fixed spin).
\item{} Large $\beta_1>0,\beta_2<0$ with $\beta_1<6\vert\beta_2\vert$ and
$\beta_1>2\vert\beta_2\vert$. In this region the vacuum is an FM configuration
on a three-dimensional cube and AF on the other direction $\mu$. We have $4$
identical possibilities to break the symmetry: one for every possible choice
of $\mu$ (more precisely $8$ if we consider also the global
$\sigma(\vec n)\to-\sigma(\vec n)$ symmetry). We call this vacuum Hyperplane
Antiferromagnetic (HPAF). The configuration is of type
$\sigma(\vec n)=(-1)^{n_\mu}\sigma_0$, where $\mu$ can be any direction.
\item{}Large $\beta_1>0,\beta_2<0$ with $\beta_1<2\vert\beta_2\vert$. In this
region the vacuum is an FM configuration on a two-dimensional plane and
AF on the other two directions. We have now six equivalent vacua.
We call this vacuum Plane Antiferromagnetic 
(PAF), and the configuration can be written as
$\sigma(\vec n)=(-1)^{n_\mu+n_\nu}\sigma_0$, where $\mu,\nu\,(\mu<\nu)$ can
 be any direction. 
\end{enumerate}

We remark that, in order to avoid undesirable (frustrating) boundary
effects for ordered phases, we work with even lattice side $L$ as
periodic boundary conditions are imposed.

Now we must define an order parameter for every phase.

For the FM phase the order parameter is the standard magnetization 
\begin{equation}
M_1=(1/V)\sum_{\vec n} \sigma(\vec n),
\end{equation}
where $V$ is the lattice volume.

In the HPAF phase we define the parameter 
$M_{2,\mu}=(1/V)\sum_{\vec n}(-1)^{n_\mu}\sigma(\vec n)$. 
$M_{2,\mu}$ will be different from
zero only in the HPAF phase, where the system becomes antiferromagnetic on the
$\mu$ direction. We have $4$ order parameters (one for every possible value
of $\mu$) and only one of them will be different from zero in the HPAF phase.
To consider only one, we define
\begin{equation}
M_2=\sqrt{\sum_{\mu}M_{2,\mu}^2}\,,
\end{equation}
that, although it is not a true order parameter, is more appropriate for
measuring on a finite lattice.

For the PAF region we first define 
$M_{3,\mu ,\nu}=(1/V)\sum_{\vec n}(-1)^{n_\mu +n_\nu}\sigma(\vec n)$. 
The same previous discussion applies, and we will work with
\begin{equation}
M_3=\sqrt{\sum_{\mu <\nu}M_{3,\mu,\nu}^2}\,.
\end{equation}

To understand the behaviour of the system, we have carried out a Mean Field
analysis and a Monte Carlo simulation.

From the Mean Field analysis we can plot a general phase diagram. We
use the standard technique, also used for gauge theories, \cite{ALESS}.
We define three different order parameters $V_1,V_2,V_3$, labelling
the FM, HPAF and PAF phases respectively.

Our Mean Field ansatz is a combination of the three possible order parameters 
\begin{equation}
V(\vec n)=V_1+(-1)^{n_t}V_2+(-1)^{n_x+n_y}V_3,
\end{equation}
with their corresponding auxiliary fields
$A(\vec n)=A_1+(-1)^{n_t}A_2+(-1)^{n_x+n_y}A_3$, where we have selected a fixed
breaking direction for the HPAF and the PAF phases.

After a standard computation, the free energy per site becomes
\begin{equation}
\begin{array}{lcr}
F  &=& -[(4\beta_1+12\beta_2)V_1^2+2\beta_1V_2^2-4\beta_2V_3^2\\
   & &+  A_1V_1+A_2V_2+A_3V_3+T],
\end{array}
\end{equation}
with $T$ defined as
\begin{equation}
\begin{array}{lcr}
T & = & (1/4)[\log\cosh(A_1+A_2+A_3)+\log\cosh(A_1+A_2-A_3)\\
  &  & +\log\cosh(A_1-A_2+A_3)+\log\cosh(A_1-A_2-A_3)].  
\end{array}
\end{equation}

By solving the saddle point equations for $F$ 
\begin{equation}
\frac{\partial F}{\partial V_i}=0, \hskip1truecm
\frac{\partial F}{\partial A_i}=0, \hskip1truecm i=1,2,3 ,
\end{equation}
we will find regions in the parameter space where the minimum of $F$ 
corresponds to different values of $V_i$.

This system of 6 equations is easily reduced to 3 equations resembling the
classical $M=\alpha\tanh(M)$ but they are coupled. By inspection one finds
some solutions:

\begin{enumerate}
\item{} $V_1=V_2=V_3=0$ is always a solution.
\item{} $V_1\ne 0$, $V_2=V_3=0$ is solution if $8\beta_1+24\beta_2>1$.
\item{} $V_2\ne 0$, $V_1=V_3=0$ if $4\beta_1>1$.
\item{} $V_3\ne 0$, $V_1=V_2=0$ if $-8\beta_2>1$.
\end{enumerate}

This divides the parameter space in several regions, some of them being
mixed. In every region one can see that those are the unique solutions, and
the deeper minimum gives the true solution for each pair ($\beta_1,\beta_2$). 
The minimum with $V_i=0 \ \forall i$ 
corresponds to the PM phase. If the minimum is at $V_1\ne 0, 
V_2=V_3=0$, we are in the FM phase. If only $V_2\ne 0$, this corresponds
to the HPAF phase and, if only $V_3\ne 0$, the phase is PAF. 
The result for the phase diagram is shown in Fig. $1$.

\begin{figure}[!t]
\psrotatefirst
\epsfig{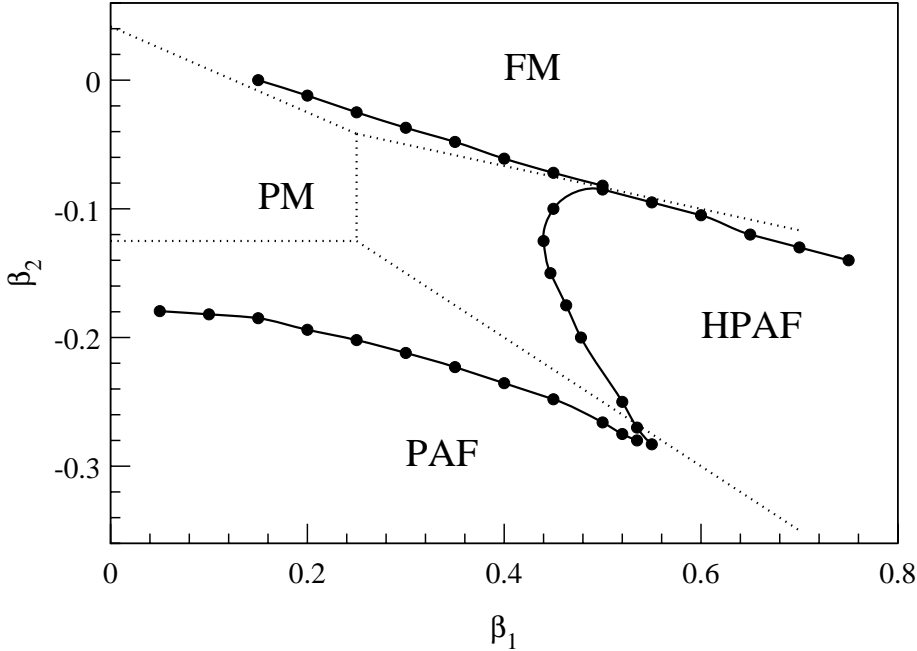}
\caption{Phase diagram obtained from Mean Field (dotted line) and Monte
Carlo (solid line and symbols, the order of the errors is of the size of the 
symbols).} 
\label{figu1.ps}
\end{figure}

The transitions separating the different phases are straight lines: 
$8\beta_1+24\beta_2=1$ between PM and FM, $4\beta_1=1$ for PM-HPAF,
$-8\beta_2=1$ for PM-PAF, $\beta_1+6\beta_2=0$ for FM-HPAF and
$\beta_1+2\beta_2=0$ for PAF-HPAF. The lines between any ordered phase and
the PM one are second order with the classical Mean Field exponents (the shape
of the equations defining these lines is identical for all of them). The lines
between two ordered phases are first order (when crossing the lines an order
parameter jumps abruptly from 0 to a positive value).

After this Mean Field approach, we have run a Heat Bath Monte Carlo
computation. 

We measure the energies, defined as 
\begin{equation} 
\langle E_1\rangle=\frac{1}{4V}\frac{\partial \log Z}{\partial
\beta_1}, \qquad
\langle E_2\rangle=\frac{1}{12V}\frac{\partial \log Z}{\partial \beta_2},
\end{equation}
and the expectation value of the order parameters, $\langle M_1\rangle,
\langle M_2\rangle,\langle M_3\rangle$.

With these quantities we have studied the global phase diagram. Using the
spectral density method \cite{FESW} and hysteresis cycles on lattices of 
size $L=8$, we have found the transition lines shown in Fig. $1$.

The line FM-PM which includes the standard Ising model point ($\beta_2=0$)
is second order with mean field exponents (from $\beta_1=0$ to some $\beta_1$
larger than the cited standard critical point).

The transition lines FM-HPAF and HPAF-PAF (see Fig. $1$)
for large values of
$\beta_1$ behave as strong first order transitions: they present metastables
states and evolve very slowly with our local Monte Carlo simulations. 
Concerning this, we address the question whether the ordered phases
are directly connected or there exists a PM  region between them. 
We could expect a PM region separating the ordered phases along 
the asymptotes $\beta_1+6\beta_2=0$ and $\beta_1+2\beta_2=0$ (narrower for
larger $\beta_1$ values) because the ground state is highly degenerate for
those values of $\beta_1, \beta_2$ (you can place an FM sublattice beside
an HPAF one without the interfase increasing the energy, if you choose the
correct orientation. You can also do this for the HPAF-PAF case).
However, from our MC
simulation it is not possible to give a conclusive answer since the width of
the hypothetical PM region decreases when increasing $\beta_1$, and for a fixed
lattice size there is a practical limit in the precision of the measures of
critical values. We have found that the lines PM-FM and PM-HPAF approach very
fast and they cannot be resolved for large $\beta_1$ values.
A similar situation is found when the PM-HPAF line comes near     
the PM-PAF one. In this case the approaching is even faster.

The transition PM-HPAF presents a clear metastability
with a large latent heat, indicating a first order           
transition.

The transition PM-PAF has a very involved behaviour. For large values
of $\beta_1$ the transition shows a large
latent heat. When we move towards smaller $\beta_1$ this latent heat
decreases, and for values $\beta_1\leq0.20$ it disappears when measured
from hysteresis cycles on lattices $L=8,\ 12,\ 16$.
This led us to think about the possibility of having a second order 
transition and study it in a more detailed way. 

On this line, we have studied three points: 
$\beta_1=0.1$, $\beta_1=0.05$ and $\beta_1=0.0$. 

At $\beta_1=0.0$ one has two decoupled lattices (one being the constituted by 
all the first neighbours of the other one). Then we have simulated separately
these lattices (we have run on $F4$ lattices). In the following, 
when we talk about the size of the lattice $L$ at the point $\beta_1=0.0$, 
we will be describing an $F4$ lattice with a number of sites $L^4/2$.  

We have simulated lattices from $L=6$ to
$L=24$. The number of Monte Carlo iterations has depended on the lattice size,
going approximately from two hundred thousand (for $L=6$) to more than one 
million (for $L=20,24$), discarding a quantity of the order of a twenty per 
cent for thermalization. 
The largest autocorrelation time found has been $O(10^3)$ for large lattices.

We have used the spectral 
density method to locate the transition points. No signal of
metastability has appeared for the smallest lattices. 
However on the largest lattices one can observe some trace of a two peak
structure. At $\beta_1=0.1$ one needs an $L=20$ lattice to begin to 
distinguish those signs
and at $\beta_1=0.05$ one has to go up to $L=24$. However at $\beta_1=0.0$ 
one can already see a two peak structure at $L=16$. 

In Fig. $2$ we show energy histograms at the three points for different 
lattice sizes. 

\begin{figure}[!t]
\epsfig{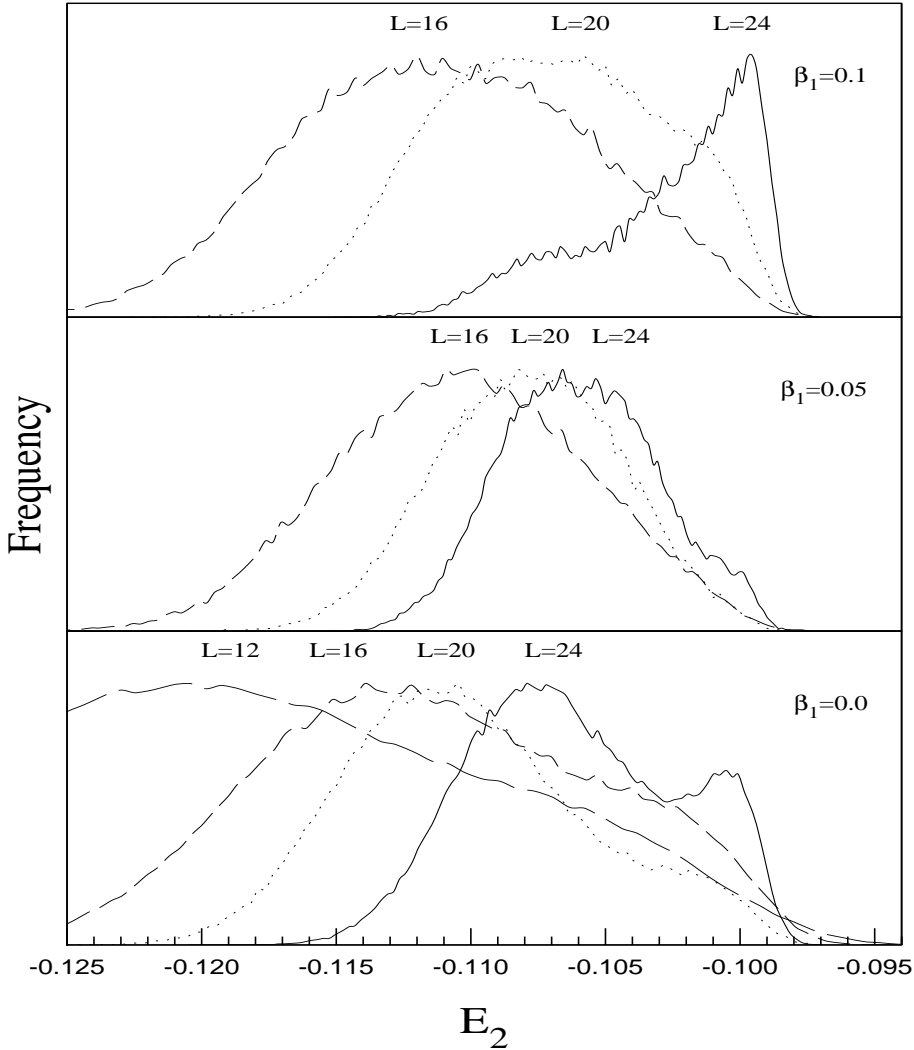}
\caption{Energy histograms for different lattice sizes in the proximities
of the three transition points studied on the line PM-PAF.}
\label{figu2.ps}
\end{figure}

Besides this direct look at the histograms, we have carried out at those 
points a Finite Size Scaling analysis 
\cite{FSSCL,FSSRE} using lattices from $L=6$ to $L=24$ in order to clarify 
the order of the transition. Despite one could not strictly talk about critical
exponents if the transition resulted to be first order, it is always possible
to calculate the exponents with which some quantities diverge with increasing
$L$ and see if they agree with those expected for a first order transition 
($\nu=1/d=0.25$, $\alpha=1$, $\gamma=1$).

We have computed the exponents $\gamma$ and $\nu$ in the way exposed below.
A good estimation of $\alpha$ has not been possible. 
We have used $\langle E_1\rangle ,\langle E_2\rangle$, 
the relevant order parameter which is the expectation value of 
$\langle M_3\rangle $ ($\langle M\rangle$ hereafter) and some of their 
derivatives. 

We start obtaining $\nu$ from the scaling law of the quantity 
\begin{equation}
\kappa_i=\frac{\partial \log \langle M\rangle}
{\partial \beta_i}=V\left(\frac{\langle ME_i\rangle}{\langle M\rangle}-
\langle E_i\rangle\right),
\end{equation}
which is
\begin{equation}
\kappa_i^{\rm max}(L)\sim L^{1/\nu}.
\end{equation}

To obtain $\gamma$ we use the
susceptibility 
\begin{equation}
\chi = V(\langle M^2\rangle-\langle M\rangle^2),
\end{equation}
with scaling law
\begin{equation}
\chi_{\rm max}(L) \sim L^{\gamma/\nu}.
\label{SCSU}
\end{equation}

The behaviour of the maximal value of $\kappa_2$ with $L$, from where we
extract $\nu$, is shown in Fig. $3$.

\begin{figure}[!t]
\psrotatefirst
\epsfig{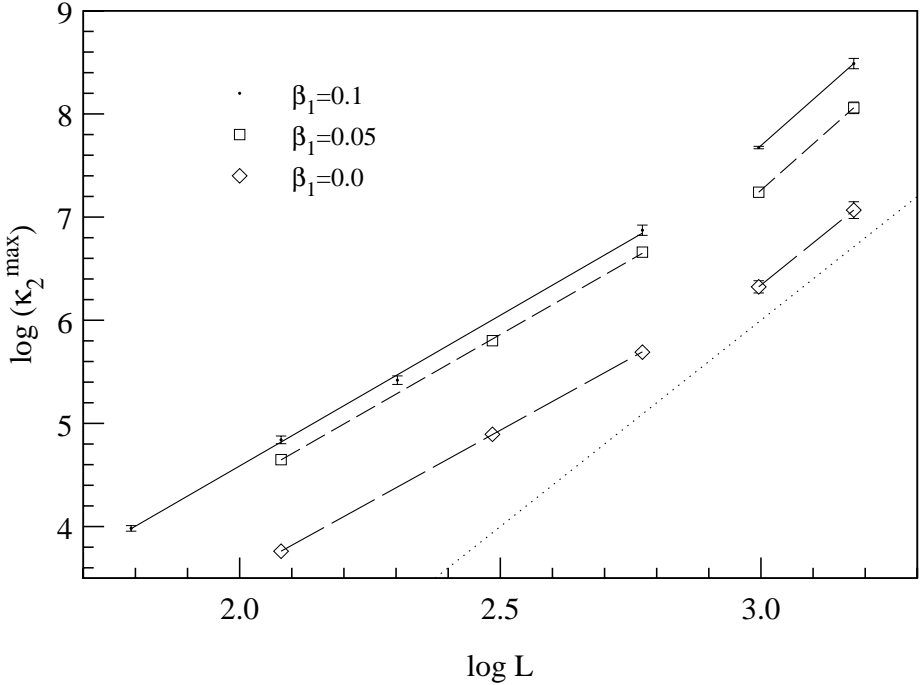}
\caption{Fits to compute $\nu$ from the maxima of $\kappa_2$. The dotted
line is for indication (its slope corresponds to $\nu=1/d$).}
\label{figu3.ps}
\end{figure}

From this figure, it is clear
that not all the $L$ values simulated are inside the asymptotic region where
one could define a correct $\nu$ exponent. Contrarily, one observes that the
exponent varies if obtained from different sets of lattice sizes. 
Notice that the behaviour for $L<20$ is surprisingly linear, with a slope 
corresponding to $\nu>1/d$. However the $L=20,24$ lattices point to a value
$\nu\simeq1/d$.



A similar behaviour is found in the computation of $\gamma/\nu$. 

The results for the exponents obtained separately from small and large 
lattices are shown in Table $1$.

\begin{table}[!t]
\begin{center}
\footnotesize
\begin{tabular}{|c|c|c|c|c|c|c|c|}
\cline{3-8}
\multicolumn{2}{c|}{\ } &
\multicolumn{2}{|c|}{$\beta_1=0.0$} &
\multicolumn{2}{|c|}{$\beta_1=0.05$} &
\multicolumn{2}{|c|}{$\beta_1=0.1$} \\ 
\hline
$d$=4 & $1^{st}$ ord &L=8,12,16 & L=20,24 & L=8,12,16 & L=20,24 & L=6,8,10,16 & L=20,24\\
\hline
\cline{1-8}
$\nu$ & 0.25 & 0.359(4) & 0.245(33) & 0.346(7) & 0.223(16) & 0.342(6) & 0.224(14)\\
\hline
$\gamma$ & 1 & 1.08(2) & 1.06(23) & 1.08(3) & 1.03(10) & 1.08(3) & 1.02(9)\\
\hline
\end{tabular}
\normalsize
\caption{Exponents obtained at the three transition points studied on the
line PM-PAF using different sets of lattice sizes. The first column refers
to the expected for a first order transition.} 
\end{center}
\end{table}


On the other hand, we tried to obtain $\alpha/\nu$ from the scaling 
of the elements of the matrix
\begin{equation}
C^{i,j}=\frac{\partial \langle E_i\rangle}{\partial \beta_j}.
\end{equation}
Its eigenvectors should be respectively orthogonal and parallel to 
the direction of the transition line. This line is, in this 
region, almost parallel to the $\beta_1$ axis and the $C^{2,2}$ element
results to be very close to the eigenvalue corresponding to the orthogonal
eigenvector, which will be the only relevant element of the matrix $C^{i,j}$
(it would not be the case if the point were a multicritical point). In short,
we studied the divergence of the $C^{2,2}$ matrix element. However we did not
get a reasonable estimation of $\alpha$. There are basically two coupled 
difficulties. Firstly the divergence of the specific heat is usually
difficult to fit. Many times one does not need just a power term in $L$ but
also a constant term or even a logarithmic or exponential one. This problem
could be solved but the added objection here is that we have very little 
data inside the asymptotic region, and the estimations we can make for 
$\alpha$ are too much poor.


A satisfactory conclusion is obtained from the behaviour of the Binder 
cumulant \cite{BINDER}
\begin{equation}
V=1-\frac{\langle E_2^4 \rangle}{3{\langle E_2^2 \rangle}^2}.
\label{BINDERCUM}
\end{equation}
The minimum of this observable approaches the value $2/3$ with increasing $L$ 
in a second 
order transition. This is not the case if the transition is first order, where
in the thermodynamical limit it takes a value depending on the position of
the two energy peaks and easily determined from (\ref{BINDERCUM}).
If these peaks are delta functions situated at $E_a$ and $E_b$, then
\begin{equation}
V_{min}=1-\frac{2({E_a}^4+{E_b}^4)}{3{({E_a}^2+{E_b}^2)}^2}.
\label{BINDERCUM2}
\end{equation}

\begin{figure}[!t]
\psrotatefirst
\epsfig{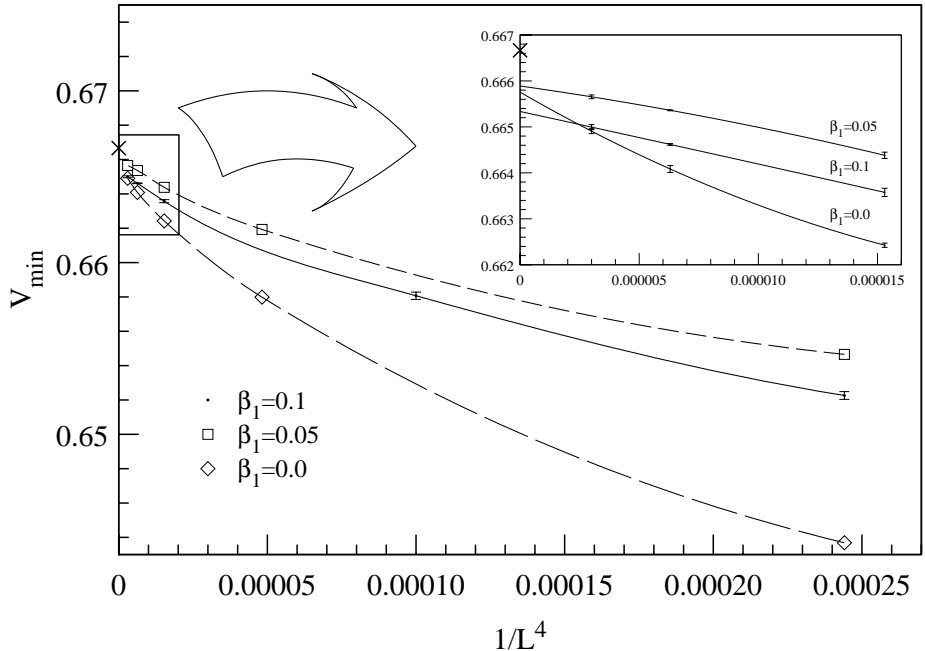}
\caption{Minimum of the Binder cumulant as a function of the lattice 
size. The cross marks the value $2/3$.}
\label{figu4.ps}
\end{figure}

 In Fig. 
$4$ it is shown how, despite its closeness to $2/3$, $V_{min}$
tends with increasing $L$ to a value slightly but significantly smaller 
than this
(approximately 0.6658) which indicates a first order transition with a latent
heat that can be estimated in the following way. Let write $E_b=E_a+\Delta$. 
Looking at the histograms (Fig. $2$) one observes that, 
approximately, one peak is situated at an energy of $-0.1$ and the latent 
heat $\Delta$ is one order of magnitude smaller. Taking that value for $E_a$
and the value $0.6658$ for $V_{min}$ in the thermodynamic limit, one obtains
from (\ref{BINDERCUM2}) a latent heat $\Delta\simeq-0.005$ which is of the 
same order as the expected from the histograms.

On the other hand, we have computed the transition points $\beta_2^c(\infty)$ 
at each $\beta_1$ value by using the expression  
\begin{equation}
\beta_2^c(\infty)=\beta_2^c(L)-AL^{-1/\nu}.
\label{betac}
\end{equation}
We have used only the largest values of $L$ and the corresponding 
exponent $\nu$.
For each observable, we have (in principle) a different $\beta_2^c(L)$
obtained as the position of the maximum derivative (with respect to $\beta_2$).
The resulting $\beta_2^c(\infty)$ is very similar for any
of them and we take the mean value.
We obtain $\beta_2^c(\infty)= -0.17904(3),\,-0.17568(8),\,-0.17459(15)$ at 
$\beta_1=0.1,0.05,0.0$ respectively. 

 As it was met in phase transitions of some other systems with frustration 
effects and non-trivial sublattice structures, 
for instance \cite{ISING2d,FCC3d}, this study has revealed to be hard
and we have needed rather large lattices to see the asymptotic behaviour,
which was masked up to a significant size ($L=24$ is a considerably large
lattice in $d=4$). On the other hand, a 
usual problem in this kind of systems is that they are difficult to 
equilibrate. However, we have simulated 64 independent lattices starting 
from different configurations and, after thermalization, they give fully 
compatible results.

In conclusion, we have described the phase diagram of the four 
dimensional Ising model with first, $\beta_1$, and second, $\beta_2$,
neighbour couplings.
In the $\beta_1>0$ semiplane,
four regions (PM, FM, HPAF and PAF) are present. The $\beta_1<0$ semiplane 
is obtained from the above by means of a trivial transformation. 
All the transition lines have resulted to be first order except that 
corresponding to the first neighbour Ising model which is second
order with Mean Field exponents.
In consequence, this simple modification to the Ising model does not seem, 
by itself, to be useful 
for finding a critical behaviour different from the classical one.






\bigskip

{\bf Acknowledgments}
We wish to thank Juan J. Ruiz-Lorenzo for useful discussions, as well as
the RTN group for the use of the RTN machine, where part 
of this  work has been done. 
Partially supported
by CICyT AEN93-0604-C03, AEN94-0218, AEN93-0776.
D.I., J.C.G. and J.M.C. are MEC Fellows and C.L.U. DGA Fellow.


\end{document}